\documentstyle[12pt]{article}
\makeatletter
\newcommand{\singlespacing}{\let\CS=\@currsize\renewcommand{\baselinestretch}
{1.0}\tiny\CS}
\newcommand{\doublespacing}{\let\CS=\@currsize\renewcommand{\baselinestretch}
{1.5}\tiny\CS}

\begin{document}

\title{Causality, Joint measurement and Tsirelson's bound}
\author{Sujit K. Choudhary, Guruprasad Kar, Samir Kunkri and Ramij Rahaman \\
Physics and Applied Mathematics Unit,\\ Indian Statistical
Institute,\\ 203 B. T. Road, Kolkata 700108,\\
India\\ {\small email : sujit\_r@isical.ac.in, gkar@isical.ac.in,
skunkri\_r@isical.ac.in, ramij\_r@isical.ac.in }}

\date{}
 \maketitle \vspace{0.5cm}

\begin{center}
{\bf Abstract}
\end{center}

{\small Tsirelson showed that $2\sqrt{2}$ is the maximum value
that CHSH expression can take for quantum-correlations [B. S.
Tsirelson {\it Lett. Math. Phys. } {\bf 4}, (1980) 93]. This bound
simply follows from the algebra of observables. Recently by
exploiting the physical structure of quantum mechanics like
unitarity and linearity, Buhrman and Massar [H. Buhrman and S.
Massar {\it Phys. Rev. A } {\bf 72}, (2005) 052103] have
established that violation of Tsirelson's bound in quantum
mechanics will imply signalling. We prove the same with the help
of realistic joint measurement in quantum mechanics and a Bell's
inequality which has been derived under the assumption of
existence of joint measurement and no signalling condition. }

 \section{Introduction}

There exists quantum-mechanical states shared between two parties
which exihibit nonlocal character. This nonlocality is quantified
by using `Bell's expression'. This is an expression which is
bounded by a certain value for `Local Hidden Variable (LHV)
models'; but can exceed this value in case of quantum
correlations. Consider for example a setting of two parties, Alice
and Bob; sharing a quantum state $\rho$ and each has a choice of
two local measurements. Alice can measure the observables $A$ and
$A^{\prime}$ whereas Bob's observables are $B$ and $B^{\prime}$.
The measured values of all the obsevables can be 1 or -1. One
relevant Bell's expression in this case is the
Clauser-Horn-Shimony-Holt (CHSH) expression \cite{chsh}. For local
hidden varriable models, this expression is bounded by 2 but in
case of entangled quantum systems,this bound can be violated. For
example, on the singlet state of two qubits there exist
observables ($A$,$A^{\prime}$,$B$,$B^{\prime}$)for which value of
the above expression is $2\sqrt{2}$.\\ In fact as shown later by
Tsirelson \cite{tsirelson} that $2\sqrt{2}$ is the maximum quantum
value of the CHSH expression.Tsirelson's bound is a simple
mathematical consequence of the axioms of quantum theory, but it
would be interesting to know that whether there is some deeper
reason why a violation greater than $2\sqrt{2}$ is unphysical. It
is known in this connection that a violation greater than
$\sqrt{\frac{32}{3}}\simeq3.27$ would imply that any communication
complexity problem can be solved using a constant amount of
communication \cite{W.van}.But this does not answer the question
that what odd would have happened for a violation just greater
than $2\sqrt{2}$.\\
Recently by exploiting the physical structure of quantum mechanics
like unitary dynamics and linearity; Buhrman and Massar
\cite{massar} have shown that exceeding Tsirelson's bound by
quantum mechanics will imply signalling in quantum mechanics. Here
we provide a simple proof of the same by exploiting nice results
of existence of joint measurement for spin along two different
directions in quantum mechanics \cite{kraus, lathi, busch, busch1,
kar} and a Bell's inequality derived under assumptions different
than that of the local-realism.

\section{Joint measurement, No signalling and Bell's Inequality}
Usually, Bell's inequality is derived under the notion of
local-realism. So, its violation by a theory will imply that the
theory is incompatible with the notion of local-realism. For
example, quantum mechanics violates it and so we conclude that
quantum theory is not local-realistic. Tsirelson observed that for
quantum mechanical states and observables, Bell's expression can
go maximum upto $2\sqrt{2}$. Here arises an interesting question
that what unphysical would have happened if quantum mechanics had
violated Tsirelson's bound. One cannot answer this question on the
basis of a Bell's inequality derived under the notion of
local-realism. Bell's inequality in this context can only tell
that quantum mechanics is not local-realistic, it cannot tell more
than this.\\

Recently Andersson et. al \cite{andersson} have derived Bell's
inequality by assuming the existence of joint measurement (not
necessarily revealing the pre-existing value) and no signalling
condition. This is not a trivial assumption. In case of classical
system it is always possible to measure two different observables
jointly, but it is not always the case with quantum systems, where
there exist noncommuting observables. At the moment, we do not
need to think about how to achieve this joint
measurement, rather we simply assume that this can be achieved.\\

In the framework of a general non-signalling  probabilistic
theory, we consider a physical system consisting of two subsystems
shared between Alice and Bob. The two observers (Alice and Bob)
have access to one subsystem each. Assume that Bob can measure two
observables $B$ or $B{^\prime}$ on his subsystem and Alice can
measure  $A$ and $A{^\prime}$ on her's. The measured values  of
all the observables can be $1$ or $-1$. We further assume that
Alice can measure the observables $A$ and $A{^\prime}$ jointly.
Let us now consider a situation where the system is always
prepared in the same state and Alice measures $A$ and $A{^\prime}$
jointly (we shall use the
subscript $J$ to denote the joint measurement) and Bob measures the observable $B$.\\
The probability that Alice will obtain
 the result $A_J = A_J^{\prime}$ can be written as
 \begin{equation}
 p(A_J = A_J^{\prime}; B) = p(A_J = A_J^{\prime} = B) + p(A_J= A_J^{\prime} = -B)
\end{equation}
As these probabilities are non-negative, hence:
\begin{equation}
p(A_J = A_J^{\prime} = B) + p(A_J = A_J^{\prime} = -B)\geq |p(A_J
= A_J^{\prime} = B) - p(A_J = A_J^{\prime}= -B|
\end{equation}
Now the term in the right hand side can be written as
\begin{equation} |p(A_J = A_J^{\prime} = B) - p(A_J = A_J^{\prime}=
-B)| = \frac{1}{2}|E(A_J, B) + E(A_J^{\prime},B)|
\end{equation}
where the correlation function $E(A,B)$is defined as : $$E(A,B)
= p(A = B) - p(A = -B)=\overline{AB}$$ \\
 The above three equations finally give us
\begin{equation}
 p(A_J = A_J^{\prime}; B) \ge \frac{1}{2}|E(A_j, B) +
 E(A_J^{\prime},B)|
\end{equation}
Similarly, if we assume that Bob  measures for the observable
$B{^\prime}$ , we will obtain
\begin{equation}
 p(A_J = -A_J^{\prime};B{^\prime}) \ge \frac{1}{2}|E(A_j, B^{\prime}) -
 E(A_J^{\prime},B^{\prime})|
\end{equation}
Adding inequalities (4) and (5) we get:
\begin{equation}
p(A_J = A_J^{\prime}; B)+p(A_J = -A_J^{\prime};B{^\prime})\ge
\frac{1}{2}[|E(A_J, B) + E(A_J^{\prime},B)|+|E(A_j, B^{\prime}) -
 E(A_J^{\prime},B^{\prime})|]
\end{equation}
Because of the no signalling constraint the probability of Alice
getting $A_J = -A_J^{\prime}$ must be independent of the fact that
Bob measured spin along $B$ or $B^{\prime}$,{\it.{i.e.}}
\begin{equation}
p(A_J = -A_J^{\prime}; B)= p(A_J = -A_J^{\prime};B{^\prime})
\end{equation}
Putting for $p(A_J = -A_J^{\prime};B{^\prime})$ from equation (7)
into inequality (6); we get:
\begin{equation}
p(A_J = A_J^{\prime}; B)+p(A_J = -A_J^{\prime}; B)\ge
\frac{1}{2}[|E(A_J, B) + E(A_J^{\prime},B)|+|E(A_j, B^{\prime}) -
 E(A_J^{\prime},B^{\prime})|]
\end{equation}
Now, noting that, $p(A_J = A_J^{\prime}; B)+p(A_J = -A_J^{\prime};
B)=1$; inequality (8), ultimately reduces to:
\begin{equation}
 |E(A_J, B) + E(A_J^{\prime}, B)| + |E(A_J, B^{\prime}) - E(A_J^{\prime},
 B^{\prime})|\le 2
 \end{equation}
One should note that the above inequality is an usual Bell's
inequality but derived under the assumptions that there exists
joint measurement and there can be no superluminal signalling. So,
violation of this inequality in a physical theory will imply that
some or all of the assumptions used in the derivation of it are
inconsistent with that particular theory. For example if joint
measurement really exists in a physical theory then violation of
this inequality will imply signalling in that physical theory.\\
It is well known that there are quantum mechanical states which
violate this inequality. Now, in this particular context of Bell's
inequality, if no-signalling is considered to be a principle, then
violation will imply that there can be no joint-measurement in
quantum-mechanics. On the other hand to address the question of
signalling in quantum mechanics with the help of  Bell's
inequality, one will have to consider a situation in quantum
mechanics where joint measurement exists. The next two sections
deal with this situation.
\\
\section{Quantum measurements} Usual quantum measurements
are projective measurements which project the initial state of a
system to one of the eigen states of the observables being
measured. For example in a measurement for spin along direction
$\hat{\alpha}$ the projectors onto the eigenstates are:
\begin{equation}
E(\pm\hat{\alpha})=\frac{1}{2}[I \pm \hat{\alpha}.\vec{\sigma}]
\end{equation}
But further progress had shown that the most general quantum
measurements are positive operator valued measures(POVM). These
generalized measurements allow us to describe any measurement that
can be performed within the limits
of quantum mechanics.\\
In this more general framework of quantum theory, the states of a
quantum system are represented by positive trace class operators.
Most general observable is represented by a collection of positive
operators $\{E_i\}$ where $0 \le E_i \le I$ for all $i$ and $\sum
E_i = I$, $I$ being an unit operator on the Hilbert space. In a
measurement for this observable for the state $\rho$ (say), the
probability of occurance of the $i$th result is given by $Tr[\rho
E_i]$. \\In the case of spin-1/2 particles, P. Busch \cite{busch,
busch1} had first introduced collection of positive operators with
the above said properties in a particular form which can be
interpreted as unsharp spin observables. This particular unsharp
observables are represented in the following form :

\begin{equation}
E_\lambda (\hat{\alpha}) = \frac{1}{2}[I + \lambda
\hat{\alpha}.\vec{\sigma}]
\end{equation}
where $0 < \lambda \le 1$ and $\hat{\alpha}$ is an unit vector.
Here $\vec{ \sigma }
   = (\hat{\sigma_{x}},\hat{\sigma_{y}},\hat{\sigma_{z}})$
denotes the usual pauli spin operator. The spectral decomposition
of $E_\lambda(\hat{\alpha})$ is given by
\begin{equation}
E_\lambda(\hat{\alpha}) = (\frac{1 + \lambda}{2})\frac{1}{2}[I +
\hat{\alpha}.\vec{\sigma}] + (\frac{1 - \lambda}{2}) \frac{1}{2}[I
- \hat{\alpha}.\vec{\sigma}]
\end{equation}

Here $\frac{1}{2}[I + \hat{\alpha}.\vec{\sigma}]$ and
$\frac{1}{2}[I - \hat{\alpha}.\vec{\sigma}]$ are the one
dimensional orthogonal spin-projection operators on
$\emph{H}^{\emph{\textbf{2}}}$. From this representation it is
clear that the POVM $\{{E_\lambda (\hat{\alpha}),E_\lambda
(\hat{-\alpha})}\}$ is a smeared version of the projective
measurement$\{\frac{1}{2}[I + \hat{\alpha}.\vec{\sigma}]
,\frac{1}{2}[I -\hat{\alpha}.\vec{\sigma}]\} $. This is the formal
sense in which the former represents unsharp spin measurement in
the direction $\hat{\alpha}$ . Noteworthy here is that for
$\lambda=1$, it represents the usual sharp (projective) spin
measurement along $\hat{\alpha}$. The eigenvalues r and u of
$E_{\lambda}(\hat{\alpha})$ where;
\\$$r=\frac{1}{2}(1 + \lambda)>\frac{1}{2}
$$and$$u = \frac{1}{2}(1 - \lambda)< \frac{1}{2}$$
are interpretated respectively as reality degree and the degree of
unsharpness of the spin property along $\hat{\alpha}$.\\ Keeping
the above interpretation for unsharp measurement in mind it is
easy to show that expectation value of an unsharply measured spin
observable with respect to an initial state $\rho$ is proportional
to the expectation value of the corresponding spin observable when
measured sharply over the same state  $\rho$, the coefficient of
proportionality being equal to the unsharp parameter (for example
$\lambda$ in this case), as :
\begin{equation}
Tr[\rho(\hat{\alpha}.\vec{\sigma})_{U} ]=(+1) Tr [\rho
E_{\lambda}(\hat{\alpha})]+(-1)Tr[\rho E_\lambda
(\hat{-\alpha})]=\lambda Tr[\rho\hat{\alpha}.\vec{\sigma}]
\end{equation}

\section{Existence of Joint measurement in Quantum mechanics}

Projective measurements are too restrictive. In the framework of
projective measurements, there are observables which cannot be
measured jointly. This distinguishing feature of quantum mechanics
is popularly known as Complementarity. Examples of complementary
observables are position and momentum observables, spin
observables in different directions etc. But in the more general
framework, it has been shown that certain complementary
observables (in standard measurement) can be measured jointly if
they are represented by a particular form of POVM (having an
interpretation in terms of unsharpness) instead of being
represented by projection operators \cite{{kraus, lathi}}.\\Joint
measurement of spin observables in different directions has been
extensively studied by P. Busch \cite {busch}. He, by exploiting
the necessary and sufficient condition for co-existence of two
{\it effects }as given by Kraus \cite{kraus}, showed that a pair
of  unsharp spin properties $E_{\lambda_1}(\hat{\alpha_{1}})$and
$E_{\lambda_2}(\hat{\alpha_{2}})$are co-existent ({\it i.e.} can
be jointly measured) if and only if:\\
\begin{equation}
|(\lambda_1\hat{\alpha_{1}}  + \lambda_2\hat{\alpha_{2}}) | +
|(\lambda_1 \hat{\alpha_{1}} - \lambda_2\hat{ \alpha_{2}} )|\le 2
\end{equation}
For $\lambda_1 = \lambda_2 = \lambda$ {\it i.e} for equal
unsharpness for both the spin properties, the condition reduces
to:
\begin{equation}
\lambda [|\hat {\alpha_{1}} + \hat{\alpha_{2}}| +
|\hat{\alpha_{1}} - \hat{\alpha_{2}}|]\leq 2
\end{equation}
The term in brackets has maximum value $2\sqrt{2}$. Hence the
coexistence condition is satisfied for all pairs of directions
$\hat{\alpha_{1}} $ and
$\hat{\alpha_{2}}$ if and only if $\lambda \leq \frac{1}{\sqrt{2}}$.\\

\section{Violation of Tsirelson bound in Quantum mechanics implies violation of
causality}
 Now we consider a situation where the system consists of two,
two level quantum systems in a state $\rho$ (say). Out of the two
observers Alice and Bob, Alice; on her subsystem, measures for the
unsharp spin observables $A_{U}$ or $A_{U}^{\prime}$ (whose joint
measurement is possible in quantum mechanics) where:
$$A_{U}=\frac{1}{2}[I +\lambda
\hat{a}.\vec{\sigma}]$$ and
$$A_{U}^{\prime}=\frac{1}{2}[I + \lambda\hat{a^{\prime}}.\vec{\sigma}]$$
We will denote the sharp counterparts of these observables by $A$
and $A^{\prime}$respectively.\\  Bob on his subsystem measures
either
$$B=\frac{1}{2}[I + \hat{b} .\vec{\sigma}]$$ or
$$B^{\prime}=\frac{1}{2}[I +
\hat{b^{\prime}}.\vec{\sigma}]$$ For these observables inequality
(9) will read as :
\begin{equation}
 |E(A_U, B) + E(A_U^{\prime}, B)| + |E(A_U, B^{\prime}) - E(A_U^{\prime},
 B^{\prime})|\le 2
\end {equation}
where $E(A_U, B)$ stands for $Tr(\rho A_U B)$; $E( A_U^{\prime},
B)$ for $Tr(\rho A_U^{\prime}B)$ and so on.\\
Now from equation (13) as $Tr(\rho A_UB) = \lambda Tr(\rho AB)$,
hence we can write $E(A_U,B) = \lambda E(A,B)$ where
$E(A,B)=Tr(\rho AB)$. Similarly $E(A_U^{\prime},B) = \lambda
E(A^{\prime},B)$ and so on. It is noteworthy here that $E(A,B),
E(A^{\prime}, B)$ etc. denote the usual quantum-mechanical
expectations.\\With the help of above analysis, inequality (16)
can be rewritten as
\begin{equation}
 \lambda[|E(A,B) + E(A^{\prime}, B)| + |E(A,B^{\prime}) - E(A^{\prime},
 B^{\prime})|]\le 2
\end{equation}
As we have seen in the previous discussion that value of $\lambda$
can go maximum up to $\frac{1}{\sqrt{2}}$ in order to make joint
measurement of spin along any two different directions possible
within quantum mechanics. Hence, for no violation of the `no
signalling condition' the term in the parentheses of inequality
(17) should be either less than or equal to $2\sqrt{2}$;
{\it{i.e}} there will be no superluminal signalling in quantum
mechanics as long as :
\begin{equation}
[|E(A,B) + E(A^{\prime}, B)| + |E(A,B^{\prime}) - E(A^{\prime},
B^{\prime})|]\le 2\sqrt2
\end{equation}
{\it{i.e}} as long as quantum correlations satisfy Tsirelson's
bound.

\section{Discussion}
In the present work, we have shown that violation of Tsirelson's
bound in quantum mechanics will result in signalling. This we have
shown with the help of (a) POVM formalism of quantum mechanics
which limits to what extent one can simultaneously measure two non
commuting observables in quantum mechanics and (b) an inequality
due to Anderson et al \cite{andersson} derived under the
assumptions of existence of joint measurement and nonexistence of
superluminal signalling in a physical theory.\\
Fortunately, the
bound on correlation function under this newer (than the original
local-realistic) assumptions and the bound on correlation function
under the assumption of local-realism, come out to be same;
\emph{i.e.} both of these assumptions lead to the same inequality
(the Bell's inequality) \cite{foot1}. \\
This new derivation of
Bell's inequality can be exploited to search out whether a theory
permits signalling or not, for it's violation in a theory will
imply that either there can be no joint measurement in that theory
or if it (joint measurement) exists,
then the theory is signalling.\\

The generalized formalism of quantum mechanics allows
joint-measurement of certain unsharp observables (not necessarily
revealing the pre existing value) provided the degree of sharpness
is sufficiently small. If such cases where joint measurement
exists are considered in quantum mechanics then violation of
Bell's inequality will imply signalling in quantum mechanics. As
its consequence, we have found that violation of Tsirelson's bound
by quantum mechanical correlation functions will result in
signalling in quantum
mechanics. \\

 Generalised observable in quantum
mechanics {\it i.e.} POVM formalism of observable captures
features of quantum mechanics in a more comprehensive way. In this
context it would be worth mentioning that Bell could construct a
Hidden Variable Theory for two dimensional quantum system by using
standard observables but it has been shown recently that if one
uses formalism of generalized observables ({\it i.e.} the POVM
formalism), then even for two dimensional quantum system,
Gleason's theorem as well as Kochen-Specker theorem hold true
\cite{busch2, cabello}. Furthermore, this formalism creates the
possibility of certain joint measurements of complementary
observables like position and momentum; spin along two different
directions etc. In particular, joint measurement of spin along
different directions are possible if standard (sharp) measurements
are replaced by their unsharp counterparts. In our case we have
used this feature of POVM formalism and it, together with a new
derivation of Bell's inequality has manifested its power by
answering an important question that the CHSH expression should be
bounded by $2\sqrt{2}$ for quantum systems to avoid superluminal
signalling in quantum mechanics.

\section{Acknowledgments}
S.K and R.R acknowledges the support by CSIR, Government of India,
New Delhi.

\end{document}